\documentclass[reprint,amsmath,nofootinbib,amssymb,aps,prl]{revtex4-2}
\usepackage{amsmath}%
\usepackage{amsfonts}%
\usepackage{amssymb}%
\usepackage[cal=cm]{mathalfa}
\usepackage{graphicx}% Include figure files
\usepackage{dcolumn}% Align table columns on decimal point
\usepackage{bm}% bold math
\usepackage{braket}
\usepackage{color}
\usepackage{bbm}
\usepackage{nicefrac}
\usepackage{float}
\usepackage{placeins}
\usepackage[dvipsnames]{xcolor}
\usepackage[utf8]{inputenc}
\usepackage[normalem]{ulem}
\usepackage{nicefrac}
\usepackage{dsfont}
\usepackage{xcolor, soul} 
\usepackage[normalem]{ulem} 
\usepackage{siunitx}
\usepackage[citebordercolor=blue]{hyperref}
\usepackage[utf8]{inputenc}
\usepackage{gensymb}
\usepackage{mathtools}
\usepackage[rightcaption]{sidecap}

\newcommand{\corrmark}[1]{%
  \textsuperscript{\hyperlink{corresponding}{#1}}%
}
\newcommand{\equalmark}{%
  \textsuperscript{\hyperlink{equalcontribution}{*}}%
}

\begin{document}

\title{A Coherent Neutron-Nucleus Hybrid System Mediated by the Strong Interaction}

\author{Niels Geerits$^{1,2,\corrmark{\dag},\equalmark}$}
\author{Kylie A. Dickerson$^{3,\corrmark{\ddag},\equalmark}$}
\author{Earl Babcock$^{3}$}
\author{Andreas Dvorak$^1$}
\author{Florian Engel$^1$}
\author{Konstantin Kholostov$^3$}
\author{Vladimir Ossovyi$^3$}
\author{Zahir Salhi$^3$}
\author{Clemens Trunner$^1$}
\author{Stephan Sponar$^{1,\corrmark{\S}}$}
\affiliation{%
$^1$Atominstitut, Technische Universit\"at Wien, Stadionallee 2, 1020 Vienna, Austria \\ $^2$Neutron Sciences Directorate, Oak Ridge National Laboratory, Oak Ridge, Tennessee 37830, USA \\
$^3$J\"ulich Centre for Neutron Science at the Heinz Maier-Leibnitz Zentrum, Forschungszentrum Jülich GmbH, Lichtenbergstr. 1, 85748 Garching, Germany}

\date{\today}
\hyphenpenalty=800\relax
\exhyphenpenalty=800\relax
\sloppy
\setlength{\parindent}{0pt}
\begin{abstract} 
A coherent ensemble of polarized helium-3 nuclei and free neutrons are coupled, forming a neutron-nucleus hybrid system. The nuclei are prepared in a spin superposition, while a magnetic field induces Zeeman splitting. The resulting nuclear energy phase is coherently transferred to the neutron, producing an intensity beat at the nuclear Zeeman frequency. Finally, the incident neutron is prepared in a spin-energy superposition state. The nuclear and neutron phases then heterodyne coherently, producing a signal at the sum or difference frequency depending on their relative helicity.
\end{abstract}

\makeatletter
\begingroup
\renewcommand{\thefootnote}{}
\renewcommand{\@makefnmark}{}
\footnotetext{\noindent
\hypertarget{equalcontribution}{%
  \textsuperscript{*} These authors contributed equally to this work.
}

\noindent
\hypertarget{corresponding}{%
  \textsuperscript{\dag}
  Corresponding author: \href{mailto:geeritsn@ornl.gov}{geeritsn@ornl.gov}\\
}
\hypertarget{corresponding}{%
  \textsuperscript{\ddag} Corresponding author:
  \href{mailto:k.dickerson@fz-juelich.de}{k.dickerson@fz-juelich.de}\\
}
\hypertarget{corresponding}{%
  \textsuperscript{\S} Corresponding author:
  \href{mailto:stephan.sponar@tuwien.ac.at}{stephan.sponar@tuwien.ac.at}
}}
\endgroup
\makeatother

\maketitle

 %“This manuscript has been authored by UT-Battelle, LLC under Contract No. DE-AC05-00OR22725 with the U.S. Department of Energy. The United States Government retains and the publisher, by accepting the article for publication, acknowledges that the United States Government retains a non-exclusive, paid-up, irrevocable, world-wide license to publish or reproduce the published form of this manuscript, or allow others to do so, for United States Government purposes. The Department of Energy will provide public access to these results of federally sponsored research in accordance with the DOE Public Access Plan (http://energy.gov/downloads/doe-public-access-plan).”

\begin{figure}
    \includegraphics[width=\textwidth/2]{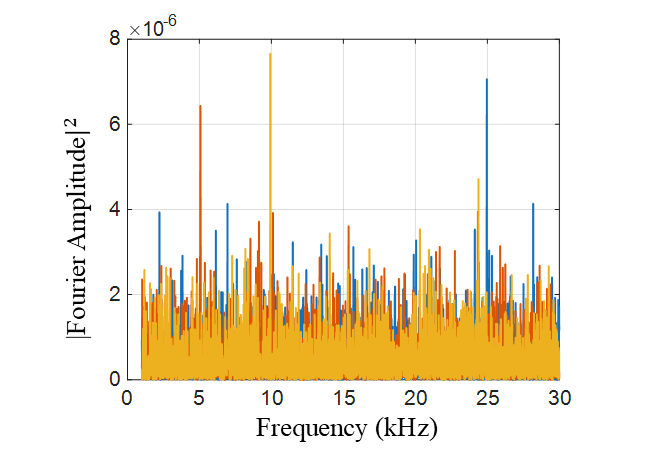}
	\caption{FFT of the neutron transmission signals for the $^3$He Zeeman frequency set to 10 kHz, with the RF frequency set to  15 kHz and the RF holding field anti-aligned with the SEOP holding field (blue), the RF frequency set to 20 kHz and the RF holding field aligned with the SEOP holding field (gold) and the RF frequency set to 15 kHz and the RF holding field aligned with the SEOP holding field (red)} 
    \label{RFsignals}
\end{figure}

\textit{Introduction - }Over the last decades the quantum mechanical nature of neutrons has been firmly established. Starting with the advent of neutron interferometry \cite{Rauch1974}, which demonstrated the wave-like nature of neutrons and the spin 1/2 4$\pi$ symmetry \cite{Werner1975,Rauch1978}, up to more recent experiments demonstrating entanglement of the neutron's quantum mechanical degrees of freedom \cite{Hasegawa2003,Hasegawa2010,Shen2020,Kuhn2021,Leiner2024} and an observation of a Leggett Garg inequality (LGI) \cite{Kreuzgruber2024}, which is a result of temporal quantum mechanical correlations. \\ 
Polarized nuclei have also been shown to exhibit quantum spin statistics through LGIs \cite{Athalye2011} and Bell tests \cite{Moussa2010}. Polarized nuclei have been studied through neutron scattering \cite{abragam1973pseudo,roubeau1974systematic,abragam1975spin}. Abragam used precessing proton spins to drive a resonant spin flip in free neutrons \cite{Abragam1972}. In addition, nuclearly polarized gases such as $^{129}$Xe, $^{131}$Xe and especially $^3$He are used extensively for neutron scattering \cite{Lu2023precise,Lu2024,Passell1966,borzakov1982features,Heil1999}. The latter is routinely utilized to polarize neutron beams, since the absorption cross section of $^3$He is highly spin dependent.
Neutrons are well isolated from the environment and could hence serve as a quantum probe. In fact, neutron scattering experiments have demonstrated entanglement in some magnetic materials \cite{Garlatti2017,Scheie2021,laurell2025witnessing}. $^3$He nuclei are also well-isolated from the environment; hence, the helium spin displays exceptional coherence, up to multiple hours with good magnetic shielding \cite{Gemmel2010}. They are therefore well suited for producing large coherent ensembles. Such polarized nuclei may serve as a test bed for quantum neutron scattering, where the neutron is used as a probe to detect quantum mechanical properties of samples.\\
In this paper we report on a series of experiments investigating a hybrid system defined by free neutrons coupled to a coherent ensemble of $^3$He nuclear spins. We argue that the strong interaction maps the phase and energy structure of the nuclear spin onto the neutrons' energy degree of freedom. In our experiments the spin of the $^3$He nuclei are prepared in a superposition state and energetically split by applying a magnetic field. Free neutrons are then passed through. The observed neutron intensity beats at the Zeeman frequency of the $^3$He nuclei. Next we add a Radio Frequency (RF) resonant spin flipper to our setup, to coherently manipulate the neutrons' energy state \cite{Sponar2008,Hasegawa2010,Shen2020,Kuhn2021,Leiner2024} and observe an intensity beating at a frequency equal to the sum or difference of the RF frequency and the $^3$He Larmor frequency, depending on the direction of the holding field. Taken together,  our experiments indicate coherent phase transfer from the nuclear ensemble to free neutrons. In addition, our theoretical analysis indicates that the final neutron state encodes the phase and energy structure of the nuclear spin state. \\
\textit{Theory -} Neutrons and polarized nuclei interact with magnetic fields through the Zeeman interaction
\begin{equation}
    \hat{V}_B = -\vec{\mu}\cdot \vec{B} = -\frac{\hbar \gamma}{2} \vec\sigma \cdot \vec B
\end{equation}
Here $\vec{\mu}$ is the particle's magnetic moment, $\vec B$ the applied magnetic field, $\gamma$ the particle's gyromagnetic ratio and $\vec\sigma$ the particle's spin operator. As a result, when exposed to a magnetic field, the particle's energy state is coupled (mode entangled) to its spin. This state can be written as
\begin{equation}\label{SpinEnergy}
        \ket{\chi}=e^{-\mathrm{i}\frac{E_\uparrow}{\hbar}t}\ket{\uparrow}+e^{-\mathrm{i}\frac{E_\downarrow}{\hbar}t}\ket{\downarrow}
\end{equation}
with $E_\downarrow-E_\uparrow=\hbar\gamma B$. \\ 
In addition to interacting with magnetic fields, neutrons and nuclei also interact with each other through the strong nuclear force, which contains a spin-independent and spin-dependent part
\begin{equation}
    V_S(\mathbf{r})=\frac{2\pi\hbar^2}{M}(b_c+\frac{1}{4}\frac{2b_i}{\sqrt{I(I+1)}}\vec\sigma_{n}\cdot\vec \sigma_{3})\delta(\mathbf{r})
\end{equation}
where $M$ is the reduced mass, $b_c$ the coherent scattering length, $b_i$ the incoherent scattering length, $I$ the nuclear spin quantum number, and $\vec\sigma_{n}$ and $\vec\sigma_{3}$ the neutron and nucleus spin operators, respectively. 
 In practice neutrons do not interact with singular nuclei, but with a bulk sample consisting of an ensemble of many nuclei; hence, the neutron sees an average potential. Nonetheless, the ensemble can be coherent, that is to say each nucleus possesses the same quantum state. The averaged neutron optical potential is given by 
 \begin{equation}\label{TruthNuke}
    \hat{V}=\frac{2\pi\hbar^2N}{m_n}(b_c+\frac{1}{4}\frac{2b_i}{\sqrt{I(I+1)}}\vec\sigma_{n}\cdot<\vec\sigma_3>)
 \end{equation}
 where $N$ is the nuclear number density in the sample, $m_n$ the neutron mass, and the nuclear spin operator $\vec\sigma_3$ is replaced by its expectation value, also known as the nuclear polarization. \\ 
     We now examine the evolution of a neutron interacting with a coherent ensemble of nuclei in the state described by equation \ref{SpinEnergy}. The nuclear polarization of this state, assuming a total nuclear polarization $P_3$ and a holding field $B_0$ along the z-axis is simply
\begin{equation}
    <\vec \sigma_3> =P_3 [\cos(\gamma_{3}B_0 \tau-\phi),-\sin(\gamma_{3}B_0 \tau-\phi), 0]^T
 \end{equation}
 Hence the spin dependent part of the potential \ref{TruthNuke} can be written using neutron spin raising/lowering operators
 \begin{equation}\label{RaisingLowering}
     \hat{V} \propto \sigma_+ e^{\mathrm{i} \gamma_3 B\tau-\mathrm{i}\phi}+\sigma_- e^{-\mathrm{i}\gamma_3 B\tau+\mathrm{i}\phi}
 \end{equation}
 indicating that the interaction can produce neutron spin-neutron energy coupling. Using first-order time dependent perturbation theory we can demonstrate that this potential drives neutron spin dependent energy transitions
 \begin{equation}\label{Golden}
    \bra{E_{f,\mp}}\hat{V}\ket{E_{i,\pm}} \propto \int d\tau \ e^{\mathrm{i}[\frac{E_i-E_f}{\hbar}\mp \mathrm{i}\gamma_3 B]\tau} = \delta(E_i-E_f\mp \hbar\gamma_3 B)
 \end{equation}
 where the $\pm$ sign refers to the spin state. We see that the transition amplitude is only non zero if $E_i-E_f = \pm \hbar \gamma_3 B$,
 indicating that our time-dependent nuclear potential drives spin-dependent energy transitions. To investigate this spin and energy transfer in more detail we will now derive the neutron wavefunction after its interaction with the coherent nuclear ensemble. The total potential seen by the neutron is the sum of the nuclear potential and the magnetic potential used to produce the state eq. \ref{SpinEnergy} in the nuclei. The Schroedinger equation is given by 
 \begin{equation}
    (-\frac{\gamma_{n}}{2}\sigma_n^z B_0+\frac{N h}{m_n}(b_c+\frac{1}{4}\frac{2b_i}{\sqrt{I(I+1)}}\vec\sigma_n\cdot<\vec \sigma_3>))\psi_n=\mathrm{i}\partial_\tau\psi_n
\end{equation}
This is the Rabi problem, which has a known solution \cite{Rabi1937,Ignatovich2003kruger}
\begin{equation}
    \psi_n(\tau)=e^{\frac{\mathrm{i}}{2}(\gamma_{3}B_0 \tau-\phi)\sigma_n^z}e^{\mathrm{i}[A+\frac{\Delta\gamma B_0}{2}\sigma_n^z+C\sigma_n^x] \tau} e^{\frac{\mathrm{i}}{2}\phi\sigma_n^z}\psi_n^0
\end{equation}
with $A=-\frac{ Nh}{m_n}b_c$, $C=-\frac{NP_3h}{2m\sqrt{I(I+1)}}b_i$, and $\Delta \gamma = \gamma_n-\gamma_3$. Note that $\phi$ is the nuclear spin phase at the time the neutron enters the sample $\phi = -\gamma_3 B_0 t$. We can ignore the global phase imprinted by the spin independent term A. Then, upon leaving the sample, the time-dependent wavefunction can be described as
\begin{equation}
    \psi_n(t)=e^{\frac{\mathrm{i}}{2}(\omega_3 \frac{d}{v}-\omega_3 t)\sigma_n^z}e^{\mathrm{i}[\frac{\Delta\omega}{2}\sigma_n^z+C\sigma_n^x] \frac{d}{v}} e^{\frac{\mathrm{i}}{2}\omega_3 t \sigma_n^z}\psi_n^0
\end{equation}
with $d$ the sample thickness, $v$ the neutron velocity, $\omega_3=\gamma_3B_0$, and $\Delta\omega=\Delta\gamma B_0$.
A few methods exist to compensate for the detuning arising from $\Delta \omega$. In this work we rely on the fact that $\Delta\gamma B_0 << C$, because both $B_0$ and $\Delta \gamma$ are small. So we will approximate the time dependent wavefunction after the sample by
\begin{equation}\label{FinalState}
    \psi_n(t)=e^{\frac{\mathrm{i}}{2}(\omega_3 \frac{d}{v}-\omega_3 t)\sigma_n^z}e^{\mathrm{i}C\sigma_n^x\frac{d}{v}} e^{\frac{\mathrm{i}}{2}\omega_3 t \sigma_n^z}\psi_n^0
\end{equation}
In the case of real $C$, the wavefunction described by equation \ref{FinalState} is almost identical to that of a neutron that traversed a resonant RF flipper, which under certain conditions can produce spin-energy entanglement \cite{Sponar2008,Hasegawa2010,Shen2020}.  \\
Our experiments use polarized $^3$He to realize the coherent nuclear ensemble. Due to strong spin-dependent absorption we will model equation \ref{FinalState}  using a large imaginary $C$, in which case the operator $e^{\mathrm{i}C\sigma_n^x\frac{d}{v}}\propto \begin{bmatrix}
    &1 &-1 \\ &-1 &1
\end{bmatrix}$. If the neutron is prepared initially in the up or down state, the final state becomes
\begin{equation}\label{InitialUp}
    \psi \propto e^{\frac{\mathrm{i}}{2}(\omega_3 \frac{d}{v}-\omega_3 t)\sigma_n^z}\begin{bmatrix}
        &\pm 1 \\ &\mp 1
    \end{bmatrix}
\end{equation}
\begin{SCfigure*}
	\includegraphics[width=13.5cm]{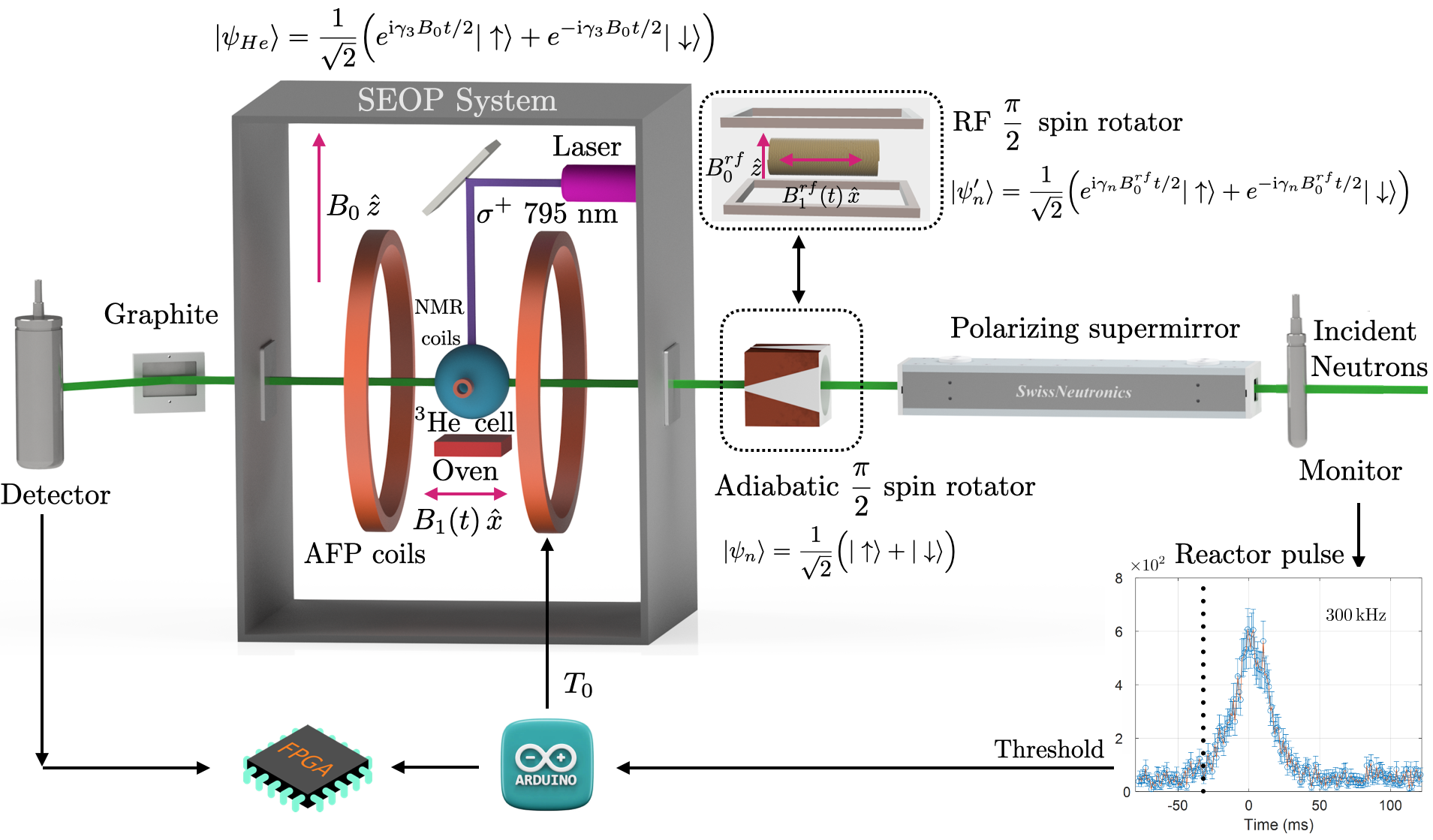}
	\caption{Diagram of setup. Neutrons are produced during a 50 ms 250\, MW reactor pulse, a unique mode of operation for TRIGA reactors. The rising edge of the pulse is registered by a monitor detector, triggering the $\pi/2$ pulse on the polarized $^3$He. The gas is polarized using a Spin-Exchange Optical Pumping (SEOP) system \cite{babcock2023situ} transferring angular momentum from circularly polarized laser light $\sigma^+$ to the nuclear spins of the $^3$He via spin-exchange collisions with the vaporized Rb. Finally neutrons are reflected from a graphite crystal and their arrival time is recorded by a detector.  }   \label{setup}
\end{SCfigure*}
which is essentially identical to the quantum state that describes each nucleus up to a constant phase factor. If, on the other hand, we start with a neutron in a spin superposition state $\psi^0_n = \frac{1}{\sqrt{2}} \begin{bmatrix}
    &e^{\mathrm{i}\varphi/2} \\ &e^{-\mathrm{i}\varphi/2}
\end{bmatrix}$, we find that each spin state is imprinted with an energy superposition, equal to the nuclear Zeeman splitting up to a constant and global phase factor
\begin{equation}
        \psi \propto e^{\frac{\mathrm{i}}{2}(\omega_3 \frac{d}{v}-\omega_3 t)\sigma_n^z}\begin{bmatrix}
        &\sin(\omega_3 t/2 +\varphi/2) \\ &-\sin(\omega_3t/2+\varphi/2)
    \end{bmatrix}
\end{equation}
or 
\begin{equation}\label{FinalFinal}
        \psi \propto e^{\frac{\mathrm{i}}{2}(\omega_3 \frac{d}{v}-\omega_3 t)\sigma_n^z} (\ket{E_+}-\ket{E_-})(\ket{\uparrow}-\ket{\downarrow})
\end{equation}
which manifests as an intensity beat
\begin{equation}\label{Int1}
    I(t) \propto \frac{1-\cos{(\omega_3 t+\varphi)}}{2}
\end{equation}
The phase $\varphi$ can in principle depend on time if the neutron spin superposition state is prepared with an RF flipper, i.e. $\varphi = \pm\omega_{RF}t + \varphi_0$. In this case the RF and nuclear signals heterodyne to form a signal at the sum or difference frequency, depending on the helicity of the respective precessions. 
\begin{equation}\label{Int2}
    I(t) \propto \frac{1-\cos{((\omega_3\pm\omega_{RF}) t+\varphi_0)}}{2}
\end{equation}
\\
\textit{Methods -} Experiments were conducted at the white beam facility of the TU Wien Atominstitut 250 kW TRIGA reactor \cite{Geerits2026}. Our experimental setup, shown in figure \ref{setup}, supports two configurations. In both configurations neutrons are polarized by double reflection from two $m=4$ supermirror polarizers before entering one of two spin control devices. In the first configuration a Hallbach array and a v-coil \cite{kraan1991adiabatic} adiabatically rotate the neutron spin into the horizontal plane, while in the second configuration a resonant RF flipper flips the neutron by $\pi/2$ into the horizontal plane and supplies a spin dependent energy kick \cite{Gahler1992,Golub1994,Ignatovich2003kruger}. The neutrons then pass through a magic box containing a polarized $^3$He cell and the SEOP system \cite{babcock2023situ,feoktystov2015kws}. Inside of the magic box, a low vertical field (2-3 G) defines the quantization axis. After neutrons pass through the system, they are monochromated by a graphite crystal. This crystal selects for a wavelength of 2.5-2.6 \AA. Finally, neutrons are detected by a $^3$He-based proportional counter. \\
The SEOP system is equipped with an RF coil, which can be used to rotate the $^3$He nuclear spin into the horizontal plane, at which point they begin to precess. Note that relative to the quantization axis, both the nuclear and the neutron spin states can be described as a superposition between up and down. In this state the coherence time $T_2$ of the nuclear spin is on the order of a few 10s of ms. As a result, one of the main challenges with this experiment is the required neutron flux to obtain sufficient statistics within the coherence window. During continuous operation at 250 kW, our setup sees only 70 neutrons every second (1-2 neutrons in our coherence window). To overcome this limitation, we use the unique pulsed operation capability of the TRIGA reactor. In this mode, the reactor can reach a peak power of 250 MW for 50 ms. Pulsed operation generates roughly 2000 neutrons per shot (i.e. 40 kHz), a sufficient number to reliably measure the beating frequency in the neutron signal arising from the transfer of the nuclear spin phase to the neutrons. Note each dataset presented here corresponds to one shot. The pulse shape is shown in figure \ref{setup}. A neutron monitor detects the incoming neutron pulse, which triggers the SEOP and detector to begin the experiment. Our two configurations allow us to measure equations \ref{Int1} and \ref{Int2} respectively, which serve as evidence for a neutron in the state \ref{FinalFinal}. 
\\
 \begin{figure}
	\includegraphics[width=\textwidth/2]{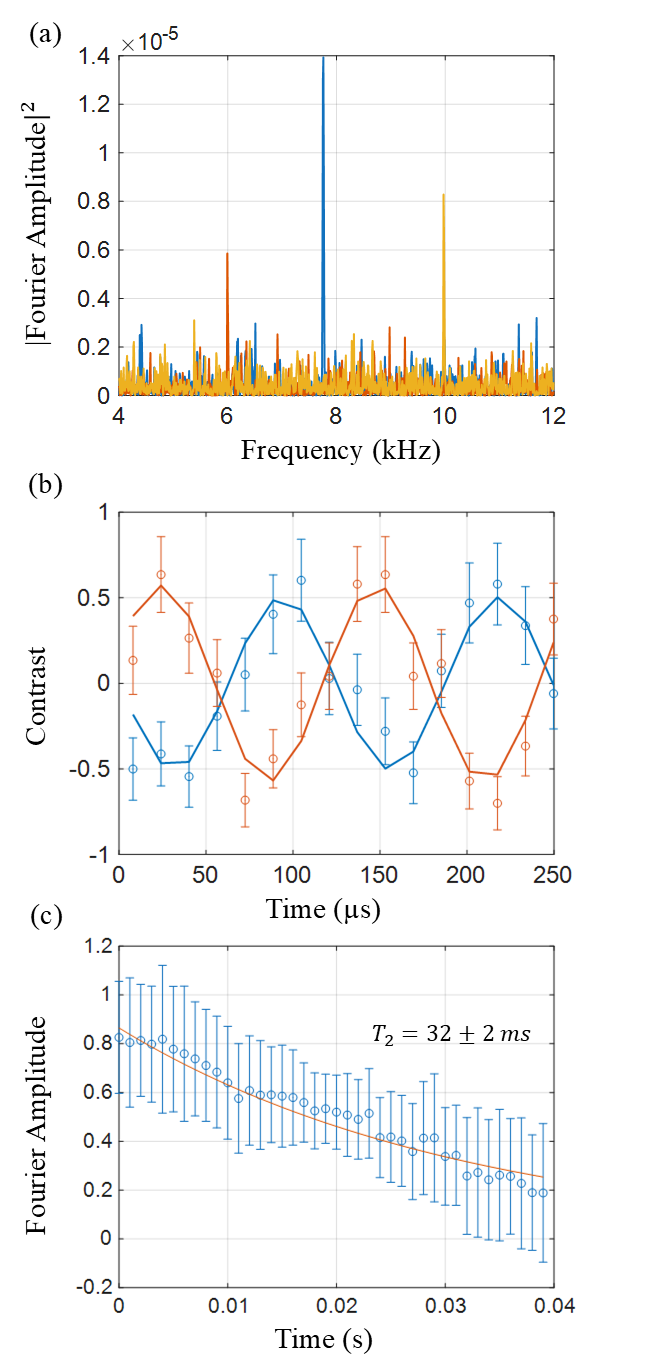}
	\caption{(a)  FFTs of raw neutron counts with the Nuclear Zeeman frequency set to 6 kHz (red), 7.75 kHz (blue) and 10 kHz (gold). (b) The time averaged neutron contrast over a 30 ms window measured as the nuclear spin precessed at a rate of 7.75 kHz (red). Afterward, the nuclear polarization was flipped by $\pi$ before starting precession (blue). (c) The Fourier amplitude at 7.75 kHz measured against time. From the decay, a $T_2$ of $32.2\pm 2$ ms can be extracted.} \label{ProcessedResults}
\end{figure}
\textit{Results -} 
In our first experiment, the nuclear precession frequency is set to 7.75 kHz. 30 ms after the start of the TRIGA pulse, the SEOP system performs a $\pi/2$ flip on the nuclear spin. 
Due to the fine time binning required to resolve the intensity beating, most time bins are empty; nonetheless, by using a Fast Fourier Transform (FFT) we can see that the intensity data beats at a frequency equal to the $^3$He precession frequency. This FFT is shown in blue in figure \ref{ProcessedResults} (a). Extraction of the time-dependent intensity beat requires that we calculate the phase of each time bin using the frequency we measured with the FFT and sum time bins of approximately equal phase. Applying this method over an area of 30 ms, we can resolve time averaged oscillations as shown in figure \ref{ProcessedResults} (b). By taking an FFT over a smaller window of the data, one can extract the average beating amplitude within that window. In addition, by shifting said time window, we can determine the oscillation amplitude against time, which in turn can be used to determine the transverse nuclear coherence time $T_2=32\pm 2$ ms. These results are shown in figure \ref{ProcessedResults} (c). To demonstrate phase control, a second measurement was conducted at the same precession frequency. Here, before starting the nuclear precession, the nuclear spin is flipped by $\pi$. As a result, we expect that the beat signal is phase shifted by $\pi$ with respect to the previous signal. Results are shown in figure \ref{ProcessedResults} (b). 
To further confirm that the Zeeman energy splitting of the $^3$He is transferred to the neutron through the neutron-nucleus spin-spin interaction, we conducted an additional measurement at holding fields of 6 kHz and 10 kHz. FFTs of the raw data are shown in figure \ref{ProcessedResults} (a). \\
Our second set of experiments are carried out in the second configuration of our setup. Here we place a resonant RF flipper between the supermirror polarizer and the SEOP box. Via a $\pi/2$ flip, the flipper provides a variable spin-dependent energy kick to the neutron. The nuclear Zeeman frequency is held constant at 10 kHz for all experiments in this configuration. We expect the RF phase to coherently add to the nuclear spin phase picked up by the neutron (ie. eq.\ref{Int2}). That is to say, we expect a new beating frequency equal to the nuclear Zeeman frequency plus or minus the RF frequency, depending on the relative directions of the RF and SEOP holding fields. Fourier transforms of the raw data are shown in figure \ref{RFsignals}. 
We find that the RF energy phase coherently adds or subtracts to/from the nuclear energy phase depending on the direction of the holding fields. When an RF frequency of 15 kHz is applied and the neutron and nuclear helicities are the same (i.e. same holding field direction), we observe a beat at the difference frequency, 5 kHz (fig. \ref{RFsignals} red). Whereas when the helicities are opposite we see a beat at the summed frequency, 25 kHz (fig. \ref{RFsignals} blue). Finally, at an RF frequency of 20 kHz and equal helicity, we observe a heterodyne signal at 10 kHz (fig. \ref{RFsignals} gold). \\
\textit{Conclusion -} We have explored a neutron-nucleus hybrid system consisting of a coherent ensemble of polarized $^3$He nuclei and single neutrons. We have shown theoretically that the phase and energy structure of the nuclear spin is mapped on to the neutrons energy degree of freedom through the spin-dependent strong interaction. The nuclear Zeeman splitting is transferred directly to the neutrons, producing intensity beating at the nuclear Larmor frequency that tracks the nuclear spin phase. When the neutron is initially prepared in an energy superposition state by means of a resonant RF flipper, the RF and nuclear precession frequencies heterodyne, producing a neutron intensity beat at either the sum or difference frequency depending on the relative helicities.  Due to this additive nature, we argue that the state produced by an RF flipper is of the same nature as the state produced through the neutron-nucleus spin-spin interaction; hence, the $^3$He produces an energy superposition state in the neutron. The neutron energy splitting equals the nuclear Zeeman energy. We conclude that the neutron state is a partial encoding of the state inhabited by the individual nuclei. \\
Our results establish this neutron-nucleus system as a coherent hybrid system in which phase and energy can be transferred and coherently combined between independently controlled subsystems, providing a new route for probing coherently prepared nuclear spin states with free neutrons.  \\
\textit{Acknowledgments -} This work was supported financially in equal parts by JCNS and the Technische Universitaet Wien. The authors acknowledge funding from the Austrian Science Fund (FWF), grant DOI: 10.55776/PAT3585625. NG acknowledges support from the Laboratory Directed Research and Development Program
LOIS \# 12612 of Oak Ridge National Laboratory, managed by UT-Battelle, LLC, for the U.S. Department of Energy. This material is based on the work supported by
the U.S. Department of Energy, Office of Science, Office of Basic Energy Sciences under Contract No. DE-AC05-00OR22725. This project has received funding from the European Union's Horizon 2020 research and innovation programme under the Marie Skłodowska-Curie grant agreement No 101034266. The authors thank Steven R. Parnell for useful discussion of the initial ideas. The authors are also grateful to the staff of the TRIGA center for making the beamtime available. Finally, the authors thank Artem Feoktysov for loaning the KWS1 $^3$He polarizer. \\
\textit{Data Availability -} The data that support the findings of this article are available from the authors upon reasonable request.

\bibliography{SpinOrbitBibliography}
\end{document}